\documentclass[showpacs,preprintnumbers,nofootinbib,
               amsmath,amssymb,superscriptaddress]{revtex4}

\newcommand{\feyn}[1]{
  \setbox0=\hbox{\ensuremath{#1}}
  \hbox to\wd0{\hbox to0pt{\hbox to\wd0{\hss/\hss}\hss}\box0}}
\newcommand{\x}{\text{x}}
\newcommand{\Det}{\text{Det}}
\newcommand{\tr}{\text{tr}}
\newcommand{\half}{{\textstyle\frac{1}{2}}}
\newcommand{\twothirds}{{\textstyle\frac{3}{2}}}
\newcommand{\diag}{\text{diag}}

\begin{document}

\title{Gauge invariant source terms in QCD}
\author{Kenji Fukushima}
\affiliation{RIKEN BNL Research Center, Brookhaven National
             Laboratory, Upton, New York 11973-5000, USA}
\begin{abstract}
 We discuss how to implement the source terms in Quantum
 Chromodynamics (QCD) respecting gauge invariance and
 noncommutativity of color charge density operators.  We start with
 decomposing the generating functional of QCD into constituents
 projected to have specified color charge density.  We demonstrate
 that such a projection leads to the gauge invariant source terms
 consisting of a naive form accompanied by the density of states which
 cancels the gauge dependence.  We then illustrate that this form is
 equivalently rewritten into a manifestly gauge invariant expression
 in terms of the Wilson line.  We confirm that noncommutativity of
 color charge density operators is fulfilled in both representations
 of the source terms.  We point out that our results are useful
 particularly to consider the problems of the quantum evolution in
 high energy QCD.
\end{abstract}
\preprint{RBRC-577}
\maketitle


\section{introduction}

     The first-principle calculations based on the fundamental theory
of the strong interaction, i.e.\ Quantum Chromodynamics (QCD), are
hindered by at least two major obstacles.  One difficulty originates
from the fact that the theory is nonlinear in gauge fields and it is
hard to find a solution even at the classical
level~\cite{Sikivie:1978sa}.  Another arises from the nonperturbative
nature, which indicates various differing aspects of QCD depending on
the context.  The most obvious manifestation of the nonperturbative
nature is, of course, that the strong coupling constant $g$ is
substantially large.  The strong coupling constant, however, runs with
the relevant momenta conveyed by gluons so that at high enough
energies the perturbative QCD (pQCD) is expected to be reliable owing
to the asymptotic freedom.

     Even when the perturbation theory seems plausible, nonlinearity
must be properly taken into account once the parton distribution
becomes large, i.e.\ the saturation effects are important in the
small-x region~\cite{Gribov:1984tu}.  Bjorken's x is the longitudinal
momentum fraction $\x=p^+/P^+$ in the infinite momentum frame where
the target hadron moves at $P^+=\infty$.  Recently powerful tools to
look into small-x wee partons have been developed from the
deliberation of this saturation picture, which leads us to an
effective theory of small-x partons described by the classical
equations of motion.  The classical description is validated because
dense partons (or gluons $\mathcal{A}_a^\mu$) at small x are expected
to behave like the classical fields created by the source $\rho^a$
that are brought in by the partons with larger
x~\cite{McLerran:1993ni,McLerran:1993ka,McLerran:1994vd}.  In the
classical model the sources are just regarded as the color charge
density carried by valence quarks inside the target hadron provided
that the target is a large nucleus.  The classical gluon fields as
strong as $\mathcal{A}_a^\mu\sim1/g$ in this picture are called the
color glass condensates.

     The classical model is the lowest-order approximation of QCD at
small x given large $\rho^a$ stemming from larger-x partons.  One can
improve the approximation by taking account of quantum corrections
around $\mathcal{A}^\mu_a$.  The theoretical framework at the one-loop
level has been well organized and the central result is now known as
the Jalilian-Marian-Iancu-McLerran-Weigert-Leonidov-Kovner (JIMWLK)
equation~\cite{Jalilian-Marian:1997jx,Jalilian-Marian:1997gr,Jalilian-Marian:1997dw,Kovner:2000pt,Iancu:2000hn,Iancu:2001ad,Ferreiro:2001qy}.
The JIMWLK equation is widely accepted, and yet, the treatment of the
source terms still leaves some subtleties, which we are addressing
here.

     First of all, the gauge invariant generalization of the source
terms is not unique.  The minimal requirements are that the source
terms are gauge invariant and are reduced to the naive (gauge variant)
form $\sim\tr[\rho A^-]$ not to affect the classical equations of
motion.  The simplest choice would be either $\sim\tr[\rho W[A^-]]$ as
adopted in Ref.~\cite{Jalilian-Marian:1997jx} or
$\sim\tr[\rho\ln W[A^-]]$ as proposed in
Ref.~\cite{Jalilian-Marian:2000ad} with anticipation from Wong's
equations, where $W[A^-]$ is the Wilson line in terms of $A^-$ in the
light-cone temporal direction.  One can also consider as many variants
as one likes which satisfy the minimal requirements.  So far,
concerning the quantum evolution at the one-loop level, all lead to
the equivalent results, but in principle, the choice should be
uniquely determined according to the approximation adopted.  We will
see later that the eikonal approximation for larger-x partons
naturally gives rise to the source terms in the form
$\sim\tr[\rho\ln W[A^-]]$ which has some intriguing properties that
$\tr[\rho W[A^-]]$ does not have.

     Second of all, the color charge is not commutative at the
operator level, that means a combination of different colors can make
another color.  In the classical model the nontrivial commutation
between color charge density is not considered since $\rho^a$ is
assumed to be large enough in the dense regime and the
noncommutativity can be ignored.  Such an approximation is no longer
valid in the dilute regime, however.  Thus it has been proposed that
the introduction of the Wess-Zumino term can handle this
issue~\cite{Kovner:2005nq,Kovner:2005aq} whose necessity is also
understandable from the gauge invariance of the source
terms~\cite{Hatta:2005wp}.  In later discussions we will clarify that
under the eikonal approximation it is the gauge variant density of
states which plays an equivalent role as the Wess-Zumino term in the
sense as discussed in Ref.~\cite{Hatta:2005wp}.  Also we will show
that the source terms expressed by $W[A^-]$ have already taken care of
noncommutativity properly.


\section{source terms}

     The QCD generating functional can be regarded as a sum of all the
contributions with distinct color charge density $\rho^a(x)$.  In this
sense we can regard the generating functional as the partition
function in grand canonical ensemble with zero chemical potential.  We
will here manage this using the projection operator which imposes a
constraint onto the functional integral.  We essentially extend the
formulation of the canonical description of QCD with respect to the
quark number \cite{Engels:1999tz,Fukushima:2002bk} to the case of
color charge density.

     We denote by $\mathcal{G}$ the Gauss operator which projects on
states with a given color charge density $\rho$ and start with
decomposing the generating functional as
\begin{equation}
 Z = \sum_{\{\rho(x)\}} Z_{\text{CE}}[\rho]
 =\sum_{\{\rho(x)\}} \int\!\mathcal{D}A\,\mathcal{D}B\;
  \mathcal{G}[A,B;\rho]\,
  e^{iS_{\text{YM}}[A]+iS_{\text{matter}}[A,B]} ,
\label{eq:decomposition}
\end{equation}
where the light-cone gauge is implicitly assumed for the gauge fixing
prescription.  We could call $Z_{\text{CE}}[\rho]$ the partition
function in the \textit{canonical ensemble} in the same sense as in
Ref.~\cite{Fukushima:2002bk}, namely, the partition function for
states with a given distribution of the color charge density.  In our
notation $B$ represents the fast-moving matter field bearing larger-x
which generates the source $\rho^a(x)$  for small-x partons.  In the
classical model of the color glass condensate which is relevant at not
very small x it is the valence quark field $\psi_a$ that $B$
represents, and at small x where quantum evolution is crucial $B$ is
the fast-moving gluon field $A^\mu_a$.  We will argue respective cases
in order below.

     From the decomposition (\ref{eq:decomposition}) we can define the
source action $S_W$ after integrating over $B$ as follows;
\begin{equation}
 Z=\sum_{\{\rho(x)\}} \int\!\mathcal{D}A\; e^{iS_{\text{YM}}[A]
  +iS_W[A;\rho]} .
\label{eq:source}
\end{equation}
All we have to do to identify $S_W[A;\rho]$ are thus to write down
$\mathcal{G}[A,B;\rho]$ and to integrate $B$ out explicitly.


\subsection{Fundamental Representation}

     Let us first consider the case that $B$ is the valence quark
field.  Here we can write the projection constraint explicitly as
follows;
\begin{equation}
 \mathcal{G}_{\text{F}}[\rho] = \prod_a\prod_x \delta \bigl[
  \rho^a(x) + g\bar{\psi}\gamma^+ t^a\psi(x) \bigr]
 = \int\!\mathcal{D}\phi\, \exp\biggl[ i\!\int d^4x\, \phi_a(x)
  \bigl\{ \rho^a(x) +g\bar{\psi}\gamma^+ t^a\psi(x)\bigr\}
  \biggr] \,,
\end{equation}
where $t^a$'s are the SU(3) algebra in the fundamental representation
which satisfy the commutation relation $[t^a,t^b]=if^{abc}t^c$ and are
normalized as $\tr\,t^a t^b=\half\delta^{ab}$.  The integration
measure associated with the $\phi$-integration is invariant under
color rotations by definition.  In the context of the canonical
ensemble $\phi_a$ is often interpreted as the imaginary chemical
potential.  The projection operator actually acts as the Laplace
(Fourier) transformation from the (imaginary) chemical potential to
the density $\rho^a$.

     The fermionic fields are now supposed to obey the antiperiodic
boundary condition with a period $2T$, which is inspired from
knowledge on the finite-temperature field theory and is known to be
suitable for describing the fermionic particle distribution.  We shall
pick up the quark propagation only in the light-cone temporal $x^+$
direction as quarks are fast moving.  This manipulation corresponds to
the eikonal approximation since neglecting spatial derivatives
$\partial_i$ means that we drop any hopping or recoil in the spatial
directions.  The action for matter is thus given by
\begin{equation}
 S_{\text{matter}}[\bar{\psi},\psi] = \int\!d^4x\;
  \bar{\psi}i\gamma^+(\partial^--igA^-_a t^a)\psi \,.
\end{equation}
If we use the notation as in the derivation of the JIMWLK
equation~\cite{Iancu:2001ad,Ferreiro:2001qy}, $A^-$ in the above
expression would be the soft gluon $\delta A^-$ actually.  There must
be its \textit{dual}, $\partial^+-ig\alpha_a t^a$, in the presence of
the classical background gauge field $\mathcal{A}^+_a=\alpha_a$ in the
covariant gauge if we consider not only the target but also the
projectile hadron.  We shall leave discussion on the projectile to
future work and neglect the latter throughout this paper.

Then the structure of the Dirac matrices becomes trivial because only
$\gamma^+$ appears and the integration over the quark fields in the
presence of the projection insertion results in the Dirac determinant,
\begin{equation}
 \mathcal{M}_{\text{F}}[A^-,\phi] = \prod_{\vec{x}}\det_{\text{Dirac}}
 \gamma^+ \Det
  \Bigl[\partial^--ig\bigl(A^-_a + \phi_a \bigr)t^a \Bigr].
\end{equation}
Here $\Det$ stands for taking the determinant in time as well as in
the color indices.  It should be noted that $\phi_a$ always appears at
the same place as $A^-_a$.  Therefore we shift the integration
variable in the following way,
\begin{equation}
 \phi_a \to \phi_a - A^-_a \,,
\label{eq:shift}
\end{equation}
in order that the determinant becomes a function of $\phi_a$ alone.

     It is quite important to note that
$\Det[\partial^--ig\phi_a t^a]=\Det[V(\partial^--ig\phi_a t^a)V^\dagger]$
indicates that $\mathcal{M}_{\text{F}}[\phi]$ is invariant under the
\textit{gauge transformation} of $\phi_a$,
\begin{equation}
 \phi_a t^a \to V\biggl(\phi_a t^a -\frac{\partial^-}{ig}\biggr)
  V^\dagger .
\label{eq:change}
\end{equation}
The determinant can be evaluated most easily if $V$ is chosen such
that $\phi_a t^a$ is static (independent of $x^+$) and diagonal in
color ($3\times3$ fundamental) space.  Once we eventually get an
expression invariant under the gauge transformation of $\phi_a$, it
can be rotated back to general $\phi_a$.  We see that a particular
choice of the periodic form,
\begin{equation}
 V^\dagger(x)= W^\phi_{x^+,-T}(\vec{x})
  \bigl[W^{\phi\dagger}_{T,-T}(\vec{x})\bigr]^{(x^++T)/2T}
  S^\dagger(\vec{x}) \,,
\label{eq:V}
\end{equation}
suffices for our purpose.  Here we defined the Wilson line at each
spatial point $\vec{x}=(x^-,\boldsymbol{x}_t)$ written as
\begin{equation}
 W^\phi_{x^+,-T}(\vec{x})
  =\mathcal{P}\exp\biggl[ig\int_{-T}^{z^+}\!\!dx^+ \phi_a(z^+,\vec{x})
  \,t^a\biggr]
\end{equation}
and $S^\dagger(\vec{x})$ is a color matrix which diagonalizes the
zero-mode.  It should be noted that $V$ must be periodic, otherwise
the antiperiodic boundary condition for $\psi$ is broken.  In
(\ref{eq:V}) the second part $[W^{\phi\dagger}_{T,-T}]^{(x^++T)/2T}$
is placed to restore the periodicity that is violated by the far left
part $W^\phi_{x^+,-T}$.  Then, substituting (\ref{eq:V}) into
(\ref{eq:change}), one acquires the transformed $\phi_a t^a$ as
\begin{equation}
 \phi_a t^a \to \diag(a_1,a_2,a_3) = \frac{1}{i2Tg}S(\ln W^\phi)
  S^\dagger \,,
\label{eq:diagonalize}
\end{equation}
where $W^\phi=W^\phi_{T,-T}$.  In general $\ln W^\phi$ is not diagonal
in color space and $S^\dagger$ is chosen so as to rotate $\ln W^\phi$
to a diagonal matrix $\diag(a_1,a_2,a_3)$ with a constraint
$a_1+a_2+a_3=0$.  We do not have to know an explicit form of
$S^\dagger$.

     Under the antiperiodic boundary condition in $x^+$ with the
period $2T$, the determinant is evaluated on the basis
$e^{i\omega_n x^+}$ with the discrete (Matsubara) frequency
$\omega_n=(2n+1)\pi/2T$.  Then by means of the formula,
$\prod_{n=0}^\infty[1-x^2/(2n+1)^2]=\cos(\pi x/2)$, we can easily
take the determinant with respect to the $x^+$ direction to have
\begin{align}
 \mathcal{M}_{\text{F}}[\phi] &= \prod_{\vec{x}}\,\det_{x^+}\bigl[
  \partial^--iga_1\bigr]^4 \det_{x^+}\bigl[\partial^--iga_2\bigr]^4
  \det_{x^+}\bigl[\partial^--iga_3\bigr]^4 \notag\\
 &=\prod_{\vec{x}}\,\bigl[\cos(Tga_1)\cos(Tga_2)\cos(Tga_3)\bigr]^4
  \notag\\
 &= \prod_{\vec{x}}\bigl[ 2+\tr W^\phi(\vec{x})
  +\tr W^{\phi\dagger}(\vec{x}) \bigr]^4 \,,
\label{eq:res_f}
\end{align}
where in the last equality we used $a_1+a_2+a_3=0$ and
\begin{equation}
 e^{i2Tga_1}+e^{i2Tga_2}+e^{i2Tga_3}
  =\tr\exp\bigl[i2Tg\;\diag(a_1,a_2,a_3)\bigr] =\tr\exp\bigl[
  S(\ln W^\phi)S^\dagger\bigr] =\tr W^\phi
\end{equation}
to make it take an obviously gauge invariant
form~\cite{Fukushima:2003fw}.  The physical meaning is now
transparent; no phase factor with the weight 2 corresponding to the
propagation of no quark and color-singlet three quarks, $W^\phi$
corresponding to the propagation of one quark, and $W^{\phi\dagger}$
the propagation of two quarks which is equivalent with that of one
antiquark.  The power 4 comes from the determinant in the Dirac
indices, namely, spin and quark-antiquark degeneracy.  We would draw
attention to the point that $\mathcal{M}_{\text{F}}$ is written only
in terms of $\tr W^\phi$, which guarantees the color commutation
relation as we will see in Sec.~\ref{sec:commutation}.

     We shall present an alternative expression of (\ref{eq:res_f})
for later convenience.  The product over $\vec{x}$ can be expanded to
result in the sum over all the charge distributions.  That is,
ignoring irrelevant overall constants and using $a_1+a_2+a_3=0$, we
can write
\begin{align}
 \mathcal{M}_{\text{F}}[\phi] &= \prod_{\vec{x}}\bigl[1+
  e^{-i2Tga_1}\bigr]^4 \bigl[1+e^{-i2Tga_2}\bigr]^4 \bigl[1+
  e^{-i2Tga_3}\bigr]^4 \notag\\
 &= \prod_{\vec{x}}\sum_{n_1(\vec{x}),n_2(\vec{x}),n_3(\vec{x})=0}^4
  \!\! _4C_{n_1(\vec{x})}\, _4C_{n_2(\vec{x})}\, _4C_{n_3(\vec{x})}\;
  e^{-i2Tg[n_1(\vec{x})a_1(\vec{x})+n_2(\vec{x})a_2(\vec{x})
  +n_3(\vec{x})a_3(\vec{x})]} \notag\\
 &= \sum_{\{\bar{\rho}(\vec{x})\}} \mathcal{W}_{\text{F}}[\bar{\rho}]
  \;e^{-\frac{2}{g}\tr[\bar{\rho} S\ln W^\phi S^\dagger]}
\label{eq:rewrite}
\end{align}
where $_4C_n=4!/(4-n)!/n!$ which is the combinatorial weight defining
the weight function $\mathcal{W}_{\text{F}}$.  From the second to the
third line we identified
$\bar{\rho}=\diag(\half gn_1,\half gn_2,\half gn_3)$ and used
(\ref{eq:diagonalize}).  The explicit form of $\mathcal{W}_{\text{F}}$
has been discussed in the random walk picture~\cite{Jeon:2004rk}, from
which not only the Gaussian form in the classical model but also the
$C$-odd terms relevant to Odderon arise~\cite{Jeon:2005cf}.  The
information of the target size should enter $\mathcal{W}_{\text{F}}$
through the measure which is necessary to convert the product over
$\vec{x}$ to the exponential of the integral over $\vec{x}$.

     In (\ref{eq:rewrite}) the summation over $\bar{\rho}(\vec{x})$
seems to be taken only with respect to the diagonal components instead
of eight color charges.  This is because $\phi_a t^a$ was chosen to
have only the diagonal components and thus the summation with respect
to off-diagonal components of $\bar{\rho}$ would, if any, result in
irrelevant constants.  Actually, even though it merely multiplies
irrelevant constants when $\phi_a t_a$ is diagonal, one should
consider that (\ref{eq:rewrite}) contains the summation with respect
to off-diagonal components as well.  This is required from the
property, $\mathcal{M}_{\text{F}}[\phi_a t^a]
=\mathcal{M}_{\text{F}}[U\phi_a t^a U^\dagger]$, which follows for any
static $U(\vec{x})$ (see (\ref{eq:change})).  Such a transformation in
(\ref{eq:rewrite}) causes the color charge rotation as
$\bar{\rho}\to U^\dagger\bar{\rho}U$.  This invariance can be restored
if the summation is symmetrically taken over all components of
$\bar{\rho}$ and moreover $\mathcal{W}_{\text{F}}$ is augmented to
have rotational symmetry in color space,
$\mathcal{W}_{\text{F}}[\bar{\rho}]=\mathcal{W}_{\text{F}}[U\bar{\rho}U^\dagger]$,
which enables us to eliminate $S(\vec{x})$ appearing in
(\ref{eq:diagonalize}) and (\ref{eq:rewrite}) by redefining
$S^\dagger\bar{\rho}S\to\bar{\rho}$.  After all we reach the general
form,
\begin{equation}
 \mathcal{M}_{\text{F}}[\phi] = \sum_{\{\bar{\rho}(\vec{x})\}}
  \mathcal{W}_{\text{F}}[\bar{\rho}]\, \exp\biggl\{ -\frac{2}{g}
  \int\!d^3x\,\tr\bigl[ \bar{\rho}(\vec{x}) \ln W^\phi(\vec{x})
  \bigr]\biggr\},
\label{eq:res_f_inv}
\end{equation}
where $\{\bar{\rho}(\vec{x})\}$ distributes not only over the diagonal
but the off-diagonal components also.


\subsection{Adjoint Representation}
\label{sec:adjoint}
 
     In the case when fast-moving partons are gluons, the projection
constraint takes a form of
\begin{equation}
 \mathcal{G}_{\text{A}}[\rho] = \prod_a\prod_x \delta \bigl[
  \rho^a(x) + D_\mu^{ab} F^{\mu+}_b \bigr]
 = \int\!\mathcal{D}\phi\, \exp\biggl[ i\!\int d^4x\, \phi_a(x)
  \bigl\{ \rho^a(x) + D_\mu^{ab} F^{\mu+}_b \bigr\} \biggr] \,.
\end{equation}
Then $\phi_a$ is always accompanied by $A^-_a$ again and we first
perform the shift (\ref{eq:shift}) as in the previous case.

     The full integration over the gluon fields is not feasible
because of the presence of nonlinear higher-order terms.  We shall
thus perform the one-loop calculation, that is, we divide the gluon
fields into fast gluons $a^\mu_a$ with large $p^+$ and slow gluons
$A^\mu_a$ with small $p^+$, and expand the Yang-Mills action in terms
of $a^\mu_a$ up to the quadratic order, and then perform the Gaussian
integration with respect to $a^\mu_a$.

     Under the eikonal approximation that $p^+$ is much larger than
other scales we shall consider in the gluon propagator only the terms
involving $\partial^+D^-=\partial^+(\partial^--ig\phi_a T^a)$ with the
SU(3) algebra $T^a$ in the adjoint representation.  We make use of the
freedom similar to (\ref{eq:change}) with adjoint matrices to make
$\phi_a T^a$ static and aligned to the Cartan subalgebra (chosen as
$T^3$ and $T^8$ which are not diagonal in the adjoint representation)
as
\begin{equation}
 \phi_a T^a \to \phi_3 T^3 + \phi_8 T^8 =i\left[ \begin{array}{cccccccc}
  0 & -a_{12} & 0 & 0 & 0 & 0 & 0 & 0 \\
  a_{12} & 0 & 0 & 0 & 0 & 0 & 0 & 0 \\
  0 & 0 & 0 & 0 & 0 & 0 & 0 & 0 \\
  0 & 0 & 0 & 0 & a_{31} & 0 & 0 & 0 \\
  0 & 0 & 0 & -a_{31} & 0 & 0 & 0 & 0 \\
  0 & 0 & 0 & 0 & 0 & 0 & -a_{23} & 0 \\
  0 & 0 & 0 & 0 & 0 & a_{23} & 0 & 0 \\
  0 & 0 & 0 & 0 & 0 & 0 & 0 & 0
 \end{array}\right] \,,
\label{eq:matrix}
\end{equation}
where we defined $a_1$, $a_2$, and $a_3=-a_1-a_2$ as $\phi_3=a_1-a_2$
and $\phi_8=\sqrt{3}(a_1+a_2)$ in accord to (\ref{eq:diagonalize}).
We further defined $a_{12}=a_1-a_2$,
$a_{23}=a_2-a_3$, and $a_{31}=a_3-a_1$.  We can regard $a_{12}$ as the
\textit{charged} gluons in the color 1-2 sector, $a_{23}$ in the 6-7
sector, and $a_{31}$ in the 4-5 sector likewise.  This matrix
structure is general also for the SU($N_c$) case, which would be
clearer in the ladder basis in color
space~\cite{KorthalsAltes:1993ca,Fukushima:2000ww}.

     After some algebra we find that the integration over the gauge
field yields the $-\half\times2=-1$ power of the determinant of the
kinetic part $\partial^+(\partial^--ig\phi_a T^a)$ in the light-cone
gauge, $A^+=0$, where 2 comes from the number of transverse gluons.
Interestingly, $\Det(\partial^--ig\phi_a T^a)$ is identical with the
Haar measure (or the Faddeev-Popov
determinant~\cite{Reinhardt:1996fs}), which eventually leads to the
Vandermonde determinant,
\begin{equation}
 \mathcal{M}_{\text{A}}^{-1}[\phi] = \prod_{\vec{x}}\Det\,\bigl(
  \partial^--ig\phi_a T^a \bigr)
 = \prod_{\vec{x}}\prod_{i<j}\;
  \Bigl|e^{i2Tga_i} - e^{i2T ga_j} \Bigr|^2
 = \prod_{\vec{x}}\sin^2(Tga_{12}) \sin^2(Tga_{23}) \sin^2(Tga_{31})
\label{eq:Haar}
\end{equation}
under the periodic boundary condition with the period $2T$.  We note
that the periodic boundary condition is appropriate to realize the
distribution of bosonic particles.  One can readily confirm
(\ref{eq:Haar}) in the same way as in (\ref{eq:res_f}) by using the
discrete frequency $\omega_n=2n\pi/2T$ for bosons and the formula,
$\prod_{n=1}^\infty[1-x^2/n^2]=\sin(\pi x)/\pi x$.  We dropped
irrelevant overall factors.  Here the zero-mode of the 3-3 and
8-8 components of (\ref{eq:matrix}) in the frequency product seems to
render the whole determinant vanishing at a first glance, but it
should be replaced by an irrelevant \textit{nonvanishing} constant
which in fact comes from the neglected spatial derivatives
$\sim\partial_i^2$.  From another point of view, one can say that the
singular zero-mode should be removed by the complete gauge fixing on
the transverse zero-modes~\cite{Lenz:1994cv}.  In any case what we
need to know is the dependence on $\phi_a T^a$ (or $a_i$) and the
zero-mode that has no effect on such dependence is safely dropped.

     It is possible to express (\ref{eq:Haar}) by the traced Wilson
line in the \textit{fundamental} representation in the same way as the
previous case of the fundamental representation, that
is~\cite{Fukushima:2003fw}
\begin{equation}
 \mathcal{M}_{\text{A}}[\phi] = \prod_{\vec{x}}\,\biggl[
  1 -\frac{2}{27}|\tr W^\phi|^2 -\frac{8}{729}\Re (\tr W^\phi)^3
  -\frac{1}{2187}|\tr W^\phi|^4 \biggl]^{-1} .
\label{eq:res_a}
\end{equation}

     It is also instructive to reexpress the results in a form similar
to (\ref{eq:res_f_inv}) that is convenient for later discussions.  By
using $\sin^2(Tga)=\half(1-\cos(2Tga))$ and expanding its inverse in
terms of $\cos(2Tga)$ one may write the determinant (not the inverse)
as
\begin{equation}
 \begin{split}
 \mathcal{M}_{\text{A}}[\phi] = & \prod_{\vec{x}}
  \sum_{n_{12}=0}^\infty\Bigl[\frac{1}{2}\bigl( e^{i2Tga_{12}}
  +e^{-i2Tga_{12}}\bigr)\Bigr]^{n_{12}} \\
 &\quad \times \sum_{n_{23}=0}^\infty \Bigl[\frac{1}{2}\bigl(
  e^{i2Tga_{23}}+e^{-i2Tga_{23}}\bigr) \Bigr]^{n_{23}}
  \sum_{n_{31}=0}^\infty \Bigl[
  \frac{1}{2}\bigl(e^{i2Tga_{31}}+e^{-i2Tga_{31}}\bigr)\Bigr]^{n_{31}}.
 \end{split}
\label{eq:rewrite_a}
\end{equation}
The intuitive meaning is now clear; there are $n_{12}$, $n_{23}$, and
$n_{31}$ gluons of $a_{12}$, $a_{23}$, and $a_{31}$ respectively at
each $\vec{x}$ and each takes either positive or negative charge with
equal probability.  The product over $\vec{x}$ can be written with an
appropriate weight function $\mathcal{W}_{\text{A}}$ in the form of
summation over the charge distributions like in (\ref{eq:rewrite}).
If the product is expanded, each term has a phase factor,
$e^{-i2Tg(m_{12}a_{12}+m_{23}a_{23}+m_{31}a_{31})}$, and we identify
$\bar{\rho}_{12}=\twothirds gm_{12}$,
$\bar{\rho}_{23}=\twothirds gm_{23}$, and
$\bar{\rho}_{31}=\twothirds gm_{31}$ to have
\begin{equation}
 \mathcal{M}_{\text{A}}[\phi] = \sum_{\{\bar{\rho}(\vec{x})\}}
  \mathcal{W}_{\text{A}}[\bar{\rho}]\, e^{-i2T\frac{2}{3}(
  \bar{\rho}_{12}a_{12}+\bar{\rho}_{23}a_{23}+\bar{\rho}_{31}a_{31})}.
\end{equation}
Here the weight function $\mathcal{W}_{\text{A}}[\bar{\rho}]$ is
suppressed exponentially for large $\bar{\rho}$ due to the higher
power of $\half$.  The following manipulation just goes like the
previous case;  we can further rewrite as
$2(\bar{\rho}_{12}a_{12}+\bar{\rho}_{23}a_{23}+\bar{\rho}_{31}a_{31})
= \frac{1}{2iTg}\tr[\bar{\rho}_{\text{ad}}S_{\text{ad}}\ln
W^\phi_{\text{ad}} S^\dagger_{\text{ad}}]$, where
$S_{\text{ad}}(\vec{x})$ and $W^\phi_{\text{ad}}$ are the adjoint
counterparts of (\ref{eq:diagonalize}) and $\bar{\rho}_{\text{ad}}$ is
the color charge density in the adjoint basis,
$\bar{\rho}_{\text{ad}}=\bar{\rho}^a T^a$.  The symmetric property
$\mathcal{M}_{\text{A}}[\phi_a T^a]=\mathcal{M}_{\text{A}}[U\phi_a T^a
U^\dagger]$ for static $U$ requires the summation over all
$\{\bar{\rho}(\vec{x})\}$ in a symmetric way here again and we finally
arrive at
\begin{equation}
 \mathcal{M}_{\text{A}}[\phi] = \sum_{\{\bar{\rho}(\vec{x})\}}
  \mathcal{W}_{\text{A}}[\bar{\rho}]\, \exp\biggl\{-\frac{1}{3g}\int\!
  d^3x\,\tr\bigl[ \bar{\rho}_{\text{ad}}(\vec{x}) \ln
  W^\phi_{\text{ad}}(\vec{x}) \bigr] \biggr\}.
\label{eq:res_a_inv}
\end{equation}


\section{density of states}

     It is not hard at least to write down a formal expression for the
source terms at the present stage.  We shall first present such an
expression and then discuss its physical implication.  The source
action is deduced from the definition (\ref{eq:source}) as
\begin{align}
 \exp\bigl\{ iS_W[A^-,\rho] \bigr\} &=
  e^{-i\int\!d^4x\, \rho^a A_a^-(x)} \int\!\mathcal{D}\phi\,
  e^{i\int\!d^4x\, \rho^a\phi_a(x)} \mathcal{M}_{\text{F},\text{A}}
  [W^\phi] \notag\\
 &= e^{-i\int\!d^4x\, \rho^a A_a^-(x)}
  \mathcal{N}[\rho] \,.
\label{eq:source_density}
\end{align}
where we defined $\mathcal{N}[\rho]$ by
\begin{equation}
 \mathcal{N}[\rho] = \mathcal{M}_{\text{F},\text{A}}\Bigl[
  \mathcal{P}e^{g\int\!dx^+\frac{\delta}{\delta\rho^a(x)}t^a}\Bigr]\,
  \delta[\rho]
\label{eq:density}
\end{equation}
using that $\phi_a(x)$ is retrieved by $-i\delta/\delta\rho^a(x)$
acting on $e^{i\int\!d^4x\,\rho^a\phi_a(x)}$.  The phase factor
involving $A^-_a$ emerges as a result of the variable shift
(\ref{eq:shift}) and, interestingly, this is the only place where
$A^-_a$ appears in our expression for the source terms.

     This structure of the ``weight function'' i.e.\ a function of
$\delta/\delta\rho$ acting on the delta-functional $\delta[\rho]$
has been known in the color dipole
picture~\cite{Iancu:2003uh,Marquet:2005hu,Hatta:2005ia}.  If the
distribution of $\bar{\rho}$ in the present formalism  allows only for
a quark-antiquark pair, the color dipole picture naturally arises.
Our expression of $\mathcal{N}[\rho]$ is not restricted to the color
dipole, however, and it contains arbitrary distribution of partons
also.

     It is interesting that $\mathcal{N}[\rho]$ has a definite
physical meaning as the density of states, namely, the density of
allowed states under the constraint imposed by a given charge
density.  That is, roughly speaking,
\begin{equation}
 \mathcal{N}[\rho] \sim \sum_{\{\bar{\rho}(\vec{x})\}}
  \mathcal{W}[\bar{\rho}]\,\delta[\rho(x) + \bar{\rho}(\vec{x})] \,.
\label{eq:abelian}
\end{equation}
This above expression is, however, not exact because of the nonlinear
nature of color charge.  Here we shall only make sure this
interpretation as the density of states in the abelian case in which
the above becomes exact.  It would be helpful to gain some feeling
in such a simple case.  The counterpart of (\ref{eq:res_f}) is simply
\begin{equation}
 \prod_{\vec{x}}\Bigl[ 1+e^{g\int\!dx^+ \frac{\delta}{\delta
  \rho(x)}} +e^{-g\int\!dx^+ \frac{\delta}{\delta\rho(x)}} \Bigr] \,
  \delta[\rho]
 = \sum_{\{\bar{\rho}(\vec{x})\}} \mathcal{W}_{\text{abelian}}
  [\bar{\rho}]\, e^{\int\!d^3x\,\bar{\rho}(\vec{x})
  \int\!dx^+\frac{\delta}{\delta\rho(x)}} \,\delta[\rho] \,,
\end{equation}
where $\mathcal{W}_{\text{abelian}}[\bar{\rho}]$ allows for
$\bar{\rho}(\vec{x})=0$ or $\bar{\rho}(\vec{x})=\pm g$ at each
$\vec{x}$.  In the nonabelian case $\int\!dx^+ \delta/\delta\rho(x)$
in the above is replace by the nonabelian analogue
$\ln W[-i(\delta/\delta\rho^a(x))t^a]$.  Then it is
straightforward to reach (\ref{eq:abelian}) by using the formula,
\begin{equation}
 e^{y\frac{d}{dx}} f(x) = f(x+y) \,.
\end{equation}
Actually the exponential of the $\rho$-derivative corresponds to the
shift operator caused by the presence of partons.  As a result of a
shift in delta-functional, one parton propagation at $\vec{x}$
supplies one unit of the contribution to color charge density.  Hence
one can regard (\ref{eq:density}) as a natural nonabelian extension of
the definition of the density of states.

     We shall shortly check the gauge invariance of $S_W[A^-,\rho]$
defined by (\ref{eq:source_density}).  Under the gauge transformation,
\begin{equation}
 A^- \to V\biggl( A^- - \frac{\partial^-}{ig} \biggr) V^\dagger,\quad
 \rho \to V\rho V^\dagger \,,
\end{equation}
the source term involving the gauge fields is not gauge invariant and
acquires an additional inhomogeneous contribution,
\begin{equation}
 -i\delta_V\! \int\!d^4x\,2\tr\bigl[\rho(x)A^-(x)\bigr]
 = -\frac{1}{g}\int\!d^4x\, 2\tr\bigl[\rho(x)
  \bigl\{V^\dagger(x)\partial^- V(x)\bigr\}\bigr] \,.
\label{eq:delta_gauge}
\end{equation}
Here we chose to use the color basis in the fundamental representation
but the same argument holds in the adjoint representation as well.
The point is that the density of states is also gauge variant because
the color charge itself is gauge variant.  To investigate the
transformation property it is convenient to go back to the first line
of (\ref{eq:source_density}) and then the transformed density of
states is read as
\begin{equation}
 \mathcal{N}[V\rho V^\dagger]
 = \int\!\mathcal{D}\phi\, e^{i\int\!d^4x\, 2\tr[V\rho V^\dagger
  \phi(x)]} \mathcal{M}_{\text{F},\text{A}}[W^\phi] \,.
\label{eq:transformed}
\end{equation}
Under the variable change (\ref{eq:change}) by the same $V$ here,
$\tr W^\phi$ is invariant and thus
$\mathcal{M}_{\text{F},\text{A}}[W^\phi]$ remains unchanged, while the
exponential factor in front of it returns to a form close to that 
before the gauge transformation with discrepancy by the inhomogeneous
part. We can immediately read the additional term as
\begin{equation}
 \mathcal{N}[V\rho V^\dagger] = \exp\biggl[+\frac{1}{g}\int\!d^4x\,
  2\tr\bigl[\rho(x)\{V^\dagger(x) \partial^- V(x)\}\bigr]\biggr]
  \mathcal{N}[\rho]
\label{eq:transform_n}
\end{equation}
which exactly cancels (\ref{eq:delta_gauge}) out.  Therefore the
source terms (\ref{eq:source_density}) are certainly gauge invariant
as a whole, though each part is not invariant.  The essence is that we
do not have to deform $A^-$ to be gauge invariant by means of the
Wilson line.  The inhomogeneous terms resulting from the gauge
transformation $V$ involve \textit{not} $A^-$ but only $V$ and
$\rho^a$.  Therefore one may as well expect such terms to be
eliminated by a factor given only in terms of $\rho^a$, the role of
which is actually played by the density of states.  In
Ref.~\cite{Hatta:2005wp} the Wess-Zumino term is also characterized as
a special solution to the transformation
property~(\ref{eq:transform_n}).  It would be an interesting future
problem to look for a relation between $\mathcal{N}[\rho]$ and the
Wess-Zumino term, which is not clear at present.

     Although $S_W[A^-,\rho]$ defined by (\ref{eq:source_density}) is
certainly gauge invariant and applicable for calculations of the
quantum fluctuation, if we define the perturbation theory around the
classical solution, we do not have to face with nonlocal interactions
associated with the Wilson line in the source terms as seen in
Refs.~\cite{Iancu:2001ad,Ferreiro:2001qy}.  This would be great
advantage in a practical sense.  We point out that this fact might
have been partially observed in the recent calculations of the quantum
evolution~\cite{Hatta:2005rn}.


\section{manifestly gauge invariant form}
\label{sec:manifest}

     It is possible to convert the source terms
(\ref{eq:source_density}) into a manifestly gauge invariant form
written in terms of the Wilson line.  We started our discussion with
the generating functional (\ref{eq:decomposition}) as a sum of all the
contributions with all $\rho^a(x)$.  We can perform the
$\rho$-summation not relying on the shift operator formula as argued
previously but instead using another formula,
\begin{equation}
 \sum_x f(x)\frac{d}{dx}\delta(x)=-\frac{d}{dx}f(x)\biggl|_{x=0} \,,
\end{equation}
and then we will acquire a different and still equivalent expression
for the source terms.  After we perform the summation over $\rho(x)$,
the source part corresponding to (\ref{eq:source_density}) becomes
\begin{equation}
 \mathcal{M}_{\text{F},\text{A}}\Bigl[\mathcal{P}e^{-g\int\!dx^+
  \frac{\delta}{\delta\rho^a(x)}t^a}\Bigr] \, e^{-i\int\!d^4x\,
  \rho^a A^-_a(x)} \biggl|_{\rho=0}
 = \mathcal{M}_{\text{F},\text{A}}\Bigl[\mathcal{P}
  e^{ig\int\!dx^+ A^-_a(x)t^a}\Bigr].
\end{equation}

     Once we recall our results (\ref{eq:res_f_inv}) and
(\ref{eq:res_a_inv}) for the explicit form of the determinant, we can
easily deduce from the definition (\ref{eq:source}) the alternative
representation of the source terms,
\begin{equation}
 \exp\bigl\{iS_W[A^-,\bar{\rho}]\bigr\} = \mathcal{W}_{\text{F}}
  [\bar{\rho}]\,\exp\biggl\{-\frac{2}{g}\int\!d^3x\,\tr\bigl[
  \bar{\rho}(\vec{x}) \ln W[A^-](\vec{x}) \bigr] \biggr\}
\label{eq:log_f}
\end{equation}
in the fundamental representation with $\bar{\rho}=\bar{\rho}^a t^a$
or
\begin{equation}
 \exp\bigl\{iS_W[A^-,\bar{\rho}]\bigr\} = \mathcal{W}_{\text{A}}
  [\bar{\rho}]\,\exp\biggl\{-\frac{1}{3g}\int\!d^3x\,\tr
  \bigl[\bar{\rho}_{\text{ad}}(\vec{x}) \ln W_{\text{ad}}[A^-]
  (\vec{x})\bigr] \biggr\}
\label{eq:log_a}
\end{equation}
in the adjoint representation with
$\bar{\rho}_{\text{ad}}=\bar{\rho}^a T^a$.  It should be noted that in
this representation the summation over $\rho^a(x)$ gives way to the
summation over $\bar{\rho}(\vec{x})$ which derives from the
integration over matter $B$.  It has a significant meaning that
$\bar{\rho}(\vec{x})$ appearing in (\ref{eq:log_f}) and
(\ref{eq:log_a}) should be distinct from $\rho(x)$.  In fact, as we
will discuss in Sec.~\ref{sec:commutation}, $\bar{\rho}(\vec{x})$ is,
so to speak, the \textit{classical} source if $\rho(x)$ is referred to
as the \textit{quantum} one.  Besides, since $\bar{\rho}(\vec{x})$
characterizes the distribution of fast-moving partons,
$\bar{\rho}(\vec{x})$ is discrete as we have seen in
(\ref{eq:rewrite}) and (\ref{eq:rewrite_a}).

     One may have thought that the fundamental and adjoint
representations should come about in accord to the quark and gluon
propagation respectively. This discrepancy is, however, only seeming
and not essential at all.  We would emphasize that the logarithmic
form is so peculiar that we should not regard it as merely one variant
of the gauge invariant generalization nor a simple alternative of
$\sim\tr[\rho W[A^-]]$.  As argued also in
Ref.~\cite{Jalilian-Marian:2000ad}, the logarithmic form does not
depend on the representation, that is,
\begin{equation}
 \frac{2}{g}\tr\bigl[\bar{\rho}\ln W[A^-]\bigr] = \frac{1}{3g}
  \tr\bigl[\bar{\rho}_{\text{ad}}\ln W_{\text{ad}}[A^-]\bigr] .
\end{equation}
In our calculations we can easily check this from
$\rho_{12}a_{12}+\rho_{23}a_{23}+\rho_{31}a_{31}
=3(\rho_1 a_1+\rho_2 a_2+\rho_3 a_3)$ in Sec.~\ref{sec:adjoint}.  From
the representation independence, it is obvious that the logarithmic
form is pure imaginary on its own.  This makes a sharp contrast to the
conventional form $\sim\tr[\rho W[A^-]]$.  If the source terms are
given by $\sim\tr[\rho W[A^-]]$ in the \textit{fundamental}
representation, the complex conjugate $\sim\tr[\rho W^\dagger[A^-]]$
would be necessary to make the source terms pure imaginary as they
should.  In the present case of the logarithmic form, on the other
hand, the source terms are always imaginary, which can be directly
seen also from
\begin{equation}
 \Bigl\{ \tr\bigl[\bar{\rho}\ln W[A^-]\bigr] \Bigr\}^\ast
  =\tr\bigl[\bar{\rho}^\dagger \ln W^\dagger[A^-]\bigr]
  = -\tr\bigl[\bar{\rho} \ln W[A^-]\bigr] \,.
\end{equation}
Here we used the hermiticity $\rho^\dagger=\rho$ and the unitarity
$W^\dagger[A^-]=W^{-1}[A^-]$.  Consequently the discrepancy between
(\ref{eq:log_f}) and (\ref{eq:log_a}) resides only in the shape of the
weight function $\mathcal{W}_{\text{F},\text{A}}[\bar{\rho}]$.


\section{commutation relation}
\label{sec:commutation}

     The color charge operators obey the commutation relation
$[\hat{\rho}^a,\hat{\rho}^b]=-igf^{abc}\hat{\rho}^c$, which can be
easily confirmed by the explicit expressions of $\hat{\rho}^a$ in
terms of quarks and gluons and the canonical quantization conditions.
From the theoretical point of view of the quantum field theory,
$[\hat{\rho}^a,\hat{\rho}^b]=-igf^{abc}\hat{\rho}^c$ is necessary for
consistency between Gauss' law and gauge invariance, that is, the
secondary constraint in the nonabelian gauge theory (i.e.\ Gauss' law)
is of the first class if we use Dirac's nomenclature for quantization
of singular systems~\cite{Dirac}.

     In the path-integral formalism the operators in the expectation
value are time-ordered from the left (large $x^+$) to the right (small
$x^+$).  Therefore the expectation value of the operator commutation
relation should be expressed as the time-ordered product;
\begin{equation}
 \langle[\hat{\rho}^a(x^+),\hat{\rho}^b(x^+)]\rangle
 =\langle\text{T}(\hat{\rho}^a(x^+)\hat{\rho}^b(x^+-\eta)
  -\hat{\rho}^b(x^+)\hat{\rho}^a(x^+-\eta))\rangle \Bigl|_{\eta\to0^+}
 =\langle\rho^a(x^+)\rho^b(x^+-\eta) -\rho^b(x^+)\rho^a(x^+-\eta)
  \rangle \Bigl|_{\eta\to0^+}.
\end{equation}
The purpose of this section is to check, in our description given in
the path-integral formalism, that the right-hand side of the above
leads to $-igf^{abc}\langle\rho^c\rangle$ as it should.

     In our formulation this noncommutative property follows from the
fact that the $\phi$-dependence in the determinant
$\mathcal{M}_{\text{F},\text{A}}[\phi]$ appears only through the
Wilson line $W^\phi$ as we have already seen in (\ref{eq:res_f}) or
(\ref{eq:res_f_inv}) and (\ref{eq:res_a}) or (\ref{eq:res_a_inv}).  It
makes no difference whether $W^\phi$ is given in the fundamental or
adjoint representation for the purpose to confirm the commutation
relation.

     Let us consider the expectation value of color charge density
evaluated with the canonical generating functional
$Z_{\text{CE}}[\rho]$.  It should be noted that $\langle\rho^a\rangle$
can be nonvanishing if it is computed with $Z_{\text{CE}}[\rho]$,
while $\langle\rho^a\rangle=0$ after the summation over $\rho$ because
of gauge invariance.  In this sense, the expectation values
$\langle\cdots\rangle$ below are to be considered as ones taken at
fixed $\rho$ in the context of the color glass condensate.

      The expectation value of color charge density is easily expressed
with the help of the identity derived from the integration of a total
derivative $\int\!d\phi(\partial/\partial\phi_a)(\cdots)=0$ inside
$Z_{\text{CE}}[\rho]$.  It immediately follows
\begin{equation}
 \langle\rho^a(x)\rangle = i\biggl\langle \frac{\delta \ln
  \mathcal{M}[W^\phi]}{\delta\phi_a(x)} \biggr\rangle
  =-g\biggl\langle \frac{\delta \ln\mathcal{M}[W^\phi]}{\delta
  W^\phi_{ij}(\vec{x})} \bigl[W^\phi_{T,x^+}(\vec{x})\, t^a
  W^\phi_{x^+,-T}(\vec{x})\bigr]_{ij} \biggr\rangle
\end{equation}
and one more derivative in $\phi^b$ from infinitesimally smaller $x^+$
yields
\begin{equation}
 \begin{split}
 \langle\rho^a(x^+)\rho^b(x^+\!-\!\eta)-\rho^b(x^+)
  \rho^a(x^+\!-\!\eta)\rangle\Bigl|_{\eta\to0^+}
 &= g^2 \biggl\langle \frac{\delta \ln\mathcal{M}[W^\phi]}{\delta
  W^\phi_{ij}(\vec{x})} \bigl[W^\phi_{T,x^+}(\vec{x})\, [t^a,t^b]
  W^\phi_{x^+,-T}(\vec{y})\bigr]_{ij} \biggr\rangle \;
  \delta^{(3)}(\vec{x}-\vec{y}) \\
 &= -igf^{abc}\bigl\langle\rho^c(x)\bigr\rangle \;
  \delta^{(3)}(\vec{x}-\vec{y}) \,.
 \end{split}
\end{equation}
This is exactly what we expect to have.  The point is that the
subtraction remains nonvanishing only when two derivatives act on the
same $W^\phi$ twice which gives rise to the commutation relation
between the algebra.

     An independent proof of the commutation relation satisfied with
the source terms taking the generic form of (\ref{eq:source_density})
has been provided in Ref.~\cite{Kovner:2005aq}.  Let us here discuss
another form (\ref{eq:log_f}) or (\ref{eq:log_a}).  The charge density
$\bar{\rho}(\vec{x})$ apparently would not satisfy the commutation
relation.  The important point is that, in view of
(\ref{eq:source_density}), $\rho^a(x)$ can be equivalently replaced by
$i\delta/\delta A^-_a(x)$ acting on $e^{iS_W[A^-]}$.  In other words,
the color charge density operator $\hat{\rho}^a(x)$ is expressed as
the derivative in $A^-_a(x)$.  (Note that $\phi_a(x)$ could be
equivalently replaced by $-i\delta/\delta\rho^a(x)$ in
Sec.~\ref{sec:manifest}.)  In this case it is not
$\bar{\rho}(\vec{x})$ but $i\delta/\delta A^-_a(x)$ acting on the
source terms which fulfills the commutation property.  In this sense
$\bar{\rho}(\vec{x})$ is classical, while $i\delta/\delta A^-_a(x)$
that is $\hat{\rho}^a(x)$ actually is quantum.  Besides, it should be
mentioned that the covariantly conserved charge is not
$\bar{\rho}(\vec{x})$ but what is inferred from
$-\delta S_W/\delta A^-_a(x)$~\cite{Jalilian-Marian:1997jx}, which
also suggests that not $\bar{\rho}(\vec{x})$ but $\rho(x)$ is the
quantum source to be considered.  It is almost trivial to check that
\begin{equation}
 \langle\rho^a(x)\rangle = -\biggl\langle \frac{\delta S_W[W[A^-]]}
  {\delta A^-_a(x)} \biggr\rangle
  =-ig\biggl\langle \frac{\delta S_W[W[A^-]}{\delta
  W[A^-]_{ij}(\vec{x})} \bigl[W_{T,x^+}(\vec{x})\, t^a W_{x^+,-T}
  (\vec{x})\bigr]_{ij} \biggr\rangle
\end{equation}
which is the covariantly conserved charge density and one more
derivative in $A^-_a$ gives
\begin{equation}
 \begin{split}
 \langle\rho^a(x^+)\rho^b(x^+\!-\!\eta)-\rho^b(x^+)
  \rho^a(x^+\!-\!\eta)\rangle\Bigl|_{\eta\to0^+}
 &= ig^2 \biggl\langle \frac{\delta S_W[W[A^-]}{\delta
  W[A^-]_{ij}(\vec{x})} \bigl[W_{T,x^+}(\vec{x})\, [t^a,t^b]
  W_{x^+,-T}(\vec{y})\bigr]_{ij} \biggr\rangle \;
  \delta^{(3)}(\vec{x}-\vec{y}) \\
 &= -igf^{abc}\bigl\langle\rho^c(x)\bigr\rangle \;
  \delta^{(3)}(\vec{x}-\vec{y}) \,.
 \end{split}
\end{equation}
From this analysis we can conclude that the noncommutative property
has already been incorporated in the framework of the JIMWLK equation
via the nonlocal source terms written in the Wilson lines even if
there is no Wess-Zumino term.


\section{summary}

     We discussed the expression of the source terms which preserves
the gauge invariance and the noncommutative nature of color charge
density operators.  It can be written, on one hand, in the form
consisting of a naive source term with the density of states.  Gauge
dependence of the naive form is canceled by the transformation
property of the density of states.  Since the interaction vertices
associated with the source terms are as simple as in the naive form in
this representation, the perturbative calculations are expected to go
effectively.  The work in this direction is now in progress.  On the
other hand, the source terms are to be expressed in a manifestly gauge
invariant form in terms of the Wilson line.  The interesting fact is
that the results are not $\sim\tr[\rho W[A^-]]$ as employed widely but
$\sim\tr[\rho\ln W[A^-]]$ which had been anticipated from Wong's
equation.  We showed how the noncommutative relation of color charge
arises in our framework, that clearly indicates that the nonlocal
interactions in the conventional JIMWLK source terms have already
handled the noncommutativity.  We believe that our new formulation
provides a useful hint to resolve some subtleties and confusions on
the source terms in high energy QCD.

\acknowledgments

     The author thanks Y.~Hatta for stimulating discussions.  He is
grateful to S.~Jeon and R.~Venugopalan for helpful comments.  He
especially thanks L.~McLerran for patient discussions and encouraging
comments.  This work was supported in part by RIKEN BNL Research
Center.

\end{document}